\documentclass[12pt]{article}
 \usepackage[cp1251]{inputenc}
 \usepackage[english]{babel}
 \usepackage{amsmath, latexsym, amssymb, bm, array, graphics, eucal,
 amsfonts,}
 \makeatletter
 \renewcommand{\@biblabel}[1]{#1.\hfill}

\makeatother

\usepackage[dvips]{graphicx}
\usepackage[flushleft,small]{caption2}

\begin{document}

 \thispagestyle{empty}
 \renewcommand{\abstractname}{\ }
 \large

 \begin{center}
\bf Friedel oscillations in quantum degenerate collisional plasma.
Screening of point charge
\end{center}\medskip

\begin{center}
  \bf  A. V. Latyshev\footnote{$avlatyshev@mail.ru$} and
  A. A. Yushkanov\footnote{$yushkanov@inbox.ru$}
\end{center}\medskip

\begin{center}
{\it Faculty of Physics and Mathematics,\\ Moscow State Regional
University,  105005,\\ Moscow, Radio str., 10--A}
\end{center}\medskip

\begin{abstract}
Research of influence of collisions on Friedel oscillations in
quantum degenerate collisional plasma ($T=0$) is carried out for the
first time. It is shown that presence of collisions in plasma leads
to exponential decreasing of amplitude and phase shift of Friedel
oscillations. In linear approximation the phase shift is equal to the
half of quantity inverse to product of Fermi's wave number by free
length path of electrons. The correct expression for longitudinal
dielectric permeability of the quantum collisional plasma found by
the authors (see arxiv:1001.3937 [math-ph] 22 Jan 2010) is used.
\medskip

\medskip

electron gas
\end{abstract}

\begin{center}\bf
  1. Introduction
\end{center}

Under classical consideration of degenerate electronic plasma
potential $V(r)$ around point charge $Q$ is described by the
classical formula of Thomas --- Fermi screening \cite{pines}
$$
V(r)=\dfrac{Q}{r}\exp(-k_{TF}r).
$$

Here $k_{TF}$ is the inverse screening radius of Thomas --- Fermi
$$
k_{TF}=\Big(\dfrac{6\pi n e^2}{\varepsilon_F}\Big)^{1/2},
$$
where $e$ is the electron charge,  $n$ is the electron concentration,
$\varepsilon_F$ is the Fermi energy.

Fridel was the first \cite{friedel1} -- \cite{friedel5} who has
found out that asymptotic (on the large distances) decreasing of
screening potential of point charge under quantum consideration of
degenerate collisionless plasma has not only monotonously
decreasing, but also oscillatory character. The reason of such
oscillations is sharp falling (to zero) of Fermi distributions for
electrons $f_F(v)$ behind Fermi's surface ,
$$
f_F(v)=\Theta(v_F-v),
$$
where $\Theta(x)$ is the Heaviside function,
$$
\Theta(x)= \left\{ \begin{array}{c}
                    1, \quad x>0, \\
                    0, \quad x<0.
                  \end{array}\right.
$$

This singularity of Fermi distribution leads to so called Kohn
singu\-la\-ri\-ties (see \cite{Kohn0} -- \cite{Grassme}). Kohn
singula\-ri\-ties are consequence of logarithmic singularities of
the longitudinal dielectric permeability  of degenerate  plasma.
Just Kohn singularities lead us to Friedel oscillations in
degenerate plasma.

In the work \cite{Grassme} the authors found out  the dependence of
Friedel's oscillations on temperature in collisionless
non-degenerate plasma. It is shown that under non-zero temperature
Friedel oscillations amplitude decreases exponentially with
distance.

In the present work the research of influence of collisions on
Friedel oscillations in quantum degenerate collisional plasma
($T=0$) is carried out. The correct expression for longitudinal
dielectric permeability of the quantum collisional plasma found by
authors (see \cite{LatYush1}) is used. It is shown that the presence
of collisions in plasma results in exponential decreasing of amplitude
and phase shift of Friedel oscillations.

In linear approximation phase shift is equal to half of the quantity
inverse to product of the Fermi wave number by free length path of
electrons. In more details, phase shift is proportional to effective
frequency of electron collisions with  plasma particles.  From here
it is clear that when the effective frequency of collisions
decreases to zero (plasma becomes collisionless), phase shift tends
to zero. It is agreed with classical result \cite{Harrison},
\cite{Landau}.

Let's emphasize that we use the expression for dielectric
longi\-tu\-dinal permeability of collisional degenerate plasma found
by us in \cite {LatYush1} . This expression is dedu\-ced on the
basis of the solution of the quantum kinetic equation for Wigner
function in coordinate space, instead of momentum space as Mernin
\cite{Mermin}. Mermin in \cite {Mermin} has received the expression
for longitudinal dielectric perme\-ability in the quantum
collisional plasma using the kinetic equation in momentum space.
However, if we use the Mermin expression for longitudinal
permeability, then exponential decreasing of amplitude of Friedel
oscillations cannot be found out. As the reason of this fact is that
the Mermin longitudinal permeability for a static case does not
depend on collision frequency.

\begin{center}
  \bf 2. Potential of point charge
\end{center}

Let us consider a point charge $Q$ in the origin of coordinates.
Then the equation describing behaviour of potential $V$ round the
point charge has the form
$$
\triangle V(\mathbf{r})=-4\pi \rho(\mathbf{r}) - 4\pi Q \delta ({\bf r}).
\eqno{(1.1)}
$$

Here $\Delta$ is the Laplace operator,  $\delta(\mathbf{r})$ is
the Dirac delta function of vector argument,
$\delta(\mathbf{r})=\delta(x)\delta(y)\delta(z)$; $\rho(\mathbf{r})$
is the density of induced charge.

We use representation of quantities in (1.1) in the form of Fourier
integrals
$$
\delta({\bf r})=\dfrac{1}{(2\pi)^3}\int e^{i{\bf kr}}d^3 k,
$$
$$
V({\bf r})=\dfrac{1}{(2\pi)^3}\int e^{i{\bf kr}}V_{\bf k}d^3 k,
$$
$$
\rho({\bf r})=\dfrac{1}{(2\pi)^3}\int e^{i{\bf kr}}\rho_{\bf k}d^3
k.
$$

Then from the equation (1.1) we receive

$$
k^2V_{\bf k}=4\pi \rho_{\bf k}+4\pi Q.
\eqno{(1.2)}
$$

Earlier we have shown \cite {LatYush1} that
$$
\rho_{\bf k}=\dfrac{{\bf k}{\bf j_k}}{\omega},\qquad
\rho_{\bf k}=\dfrac{ \sigma_l({\bf k}){\bf kE_k}}{\omega}=
-\dfrac{i\sigma_l({\bf k}) k^2}{\omega}V_{\bf k}.
$$

Here $\sigma_l(\mathbf{k})$ is the longitudinel electric conductivity
of quantum plasma  found in work \cite{LatYush1}.

Substituting this relation into (1.2) we receive
$$
k^2\Big(1+\dfrac{4\pi i\sigma_l({\bf k}) }{\omega}\Big)V_{\bf k}=4\pi
Q,
$$
or
$$
k^2\varepsilon_l({\bf k})V_{\bf k}=4\pi Q,
$$
where $\varepsilon_l(\mathbf{k})$ is the longitudinal dielectric
permeability of quantum plasma,
$$
\varepsilon_l(\mathbf{k})=1+\dfrac{4\pi i\sigma_l({\bf k})
}{\omega}.
$$

Thus spectral density of electric potential is expressed in terms of
the quantity of charge
$$
V_{\bf k}=\dfrac{4\pi Q}{k^2\varepsilon_l({ k})}.
\eqno{(1.3)}
$$

When deducing (1.3) we used the fact that dielectric permeability
depends only on the module ${\bf k}$, $\varepsilon_l({\bf k}) =
\varepsilon_l({k})$. It is obvious  that the following relation
relation $V({\bf r})=V(r)$ also takes place.

Then for the quantity $V( r)$ we have
$$
V({r})=\dfrac{4\pi Q}{(2\pi)^3}
\int \dfrac{ e^{i{\bf kr}}}{k^2\varepsilon_l({k})}d^3k=
\dfrac{4\pi Q}{(2\pi)^2}\int
\dfrac{e^{ikr\cos\theta}}{\varepsilon_l(k)}\sin\theta d\theta
dk.
$$

Integrating by the angular variable $\theta$ we receive
$$
V(r)=\dfrac{1}{r}\cdot\dfrac{2Q}{\pi}\int\limits_0^\infty
\dfrac{\sin (kr)}{\varepsilon_l(k)k}dk.
\eqno{(1.4)}
$$

The formula (1.4) corresponds to the formula received in the
monograph of Harrison (see \cite{Harrison}, p. 334) if we present
$$
V_q^0=\dfrac{Q}{4\pi q^2}.
$$

Let's consider asymptotic behaviour of potential $V(r)$, $r\to\infty
$. It is possible to consider electric field to be weak enough in
this area. Therefore linear approximation is applicable here. Fridel
oscsillations appear because of logarithmic singularities of
dielectric permeability.

After double integration of expression (1.4) by parts we find
\cite {Harrison}
$$
V(r)=\dfrac{1}{r^3}\cdot\dfrac{2Q}{\pi}\int\limits_0^\infty
\dfrac{\sin (rk)\varepsilon_l''(k)}{k\varepsilon_l^2(k)}dk+
O(e^{-k_{TF}r}),
$$
or, believing $k=qk_F$,
$$
V(r)=\dfrac{1}{r^3}\cdot\dfrac{2Q}{k_F^2\pi}\int\limits_0^\infty
\dfrac{\sin (rk_Fq)\varepsilon_l''(q)}{\varepsilon_l^2(q)}dq+
O(e^{-k_{TF}r}).
\eqno{(1.5)}
$$

In (1.5) we have left in explicit form only that term which results in
oscillations of potential. Other item designated as
$O(e^{-k_{TF}r})$
quickly decreases in exponential way, on the Thomas --- Fermi
radius it decays with distance.

Integrals of such kind as (1.5) usually tend to zero at
tendency of distance $r$ to infinity because of fast oscillations
$\sin rk$. However as it will be seen the second derivative
of $\varepsilon_l''(k)$ contains
singular Cauchy kernel that gives the non-zero contribution to integral.
Friedel oscillations are caused by logarithmic singularity of
dielectric permeability.

Function standing under the sign of integral in (1.5) is even, therefore we can expand the integration
onto whole real axis:
$$
V(r)=\dfrac{1}{r^3}\cdot\dfrac{Q}{\pi}\int\limits_{-\infty}^\infty
\dfrac{\sin (rk_Fq)\varepsilon_l''(q)}{q\varepsilon_l^2(q)}dq+
O(e^{-k_{TF}r}),\qquad r\to \infty.
\eqno{(1.6)}
$$

Let us recall that under positive values of $q>0$ the quantity $q$ equals to
$q=\dfrac{|\mathbf{k}|}{k_F}$, and under negative values the quantity $q$ has not any physical meaning.

\begin{center}
\bf 3. Friedel oscillations
\end{center}

In the expression (1.6) there is an expression of the longitudinal dielectric
permeability of quantum degenerate collisional plasma
$\varepsilon_l(q)$. According to the work \cite {LatYush1} this expression
has the following form for
statical case $\omega=0$ considered here
$$
\varepsilon_l(q)=1+\dfrac{3x_p^2}{2q^2}\dfrac{1-g_+(q)+g_-(q)}
{1-iyg_0(q)}.
\eqno{(2.1)}
$$

In (2.1)the following dimensionless parameters are introduced
$$
x_p=\dfrac{\omega_p}{k_Fv_F}, 
\qquad y=\dfrac{\nu}{k_Fv_F}=\dfrac{1}{lk_F},\quad l=v_F\tau,
$$
where $\omega_p$ is the plasma (Langmuir) frequency, $k_F=p_F/\hbar$
is the Fermi wave number, $p_F=mv_F$ is the Fermi momentum, $v_F$
is the Fermi velocity, $m$ is the electron mass, $\nu$ is the
effective electron frequency, $l=v_F/\nu$ is the electron free length
path, $\tau=1/\nu$ is the time of electron free length
path.

These functions $g_0(q), g_+(q), g_-(q)$ in statical limit  ($\omega\to 0$)
have the following form
$$
g_0(q)=\dfrac{1}{2q}\ln\dfrac{iy+q}{iy-q},
$$
$$
g_+(q)=\dfrac{1}{8q^3}\Big[(q^2+2iy)^2-4q^2\Big]\ln\dfrac{q^2+2q+2iy}
{q^2-2q+2iy},
$$
$$
g_-(q)=\dfrac{1}{8q^3}\Big[(q^2-2iy)^2-4q^2\Big]\ln\dfrac{q^2-2q-2iy}
{q^2+2q-2iy}.
$$

Let's find roots of the equations
$$
q^2\pm 2q\pm2iy=0.
$$

We have
$$
q_{1,2}=\mp 1\pm \sqrt{1\pm2iy}.
\eqno{(2.2)}
$$

Further let's consider the quantity $y$ as small parameter and designate
it through $\varepsilon$
$$
\varepsilon\equiv y=\dfrac{\nu}{k_Fv_F}=\dfrac{\nu\hbar}{mv_F^2}=
\dfrac{\nu \hbar}{2\varepsilon_F}.
$$

In linear approximation by $\varepsilon$ for roots of (2.2) we receive
$$
q_{1,2}=\mp 1\pm(1\pm i\varepsilon).
$$

We present the functions $g_+(q)$ and $g_-(q)$ in the following form
$$
g_+(q)=\dfrac{q^2+\varepsilon^2}{8q^3}(q+2-i\varepsilon)(q-2+i\varepsilon)
\ln\dfrac{(q+i\varepsilon)(q+2-i\varepsilon)}{(q-i\varepsilon)
(q-2+i\varepsilon)},
$$
$$
g_-(q)=\dfrac{q^2+\varepsilon^2}{8q^3}(q-2-i\varepsilon)(q+2+i\varepsilon)
\ln\dfrac{(q+i\varepsilon)(q-2-i\varepsilon)}{(q-i\varepsilon)
(q+2+i\varepsilon)},
$$
or, in linear approximation, rejecting terms proportional to
$\varepsilon^2$, we have

$$
g_+(q)=\dfrac{1}{8q}g_1(q),\qquad g_-(q)=\dfrac{1}{8q}g_2(q),
$$
where
$$
g_1(q)=(q-2+i\varepsilon)f_{+-}(q)-(q+2-i\varepsilon)f_{-+}(q)
$$
$$
g_2(q)=(q+2+i\varepsilon)f_{--}(q)-(q-2-i\varepsilon)f_{++}(q).
$$

Here the functions containing Kohn singularities are introduced
$$
f_{++}(q)=(q+2+ i\varepsilon)\ln(q+2+ i\varepsilon),
$$
$$
f_{--}(q)=(q-2- i\varepsilon)\ln(q-2- i\varepsilon),
$$
$$
f_{+-}(q)=(q+2- i\varepsilon)\ln(q+2- i\varepsilon),
$$
$$
f_{-+}(q)=(q-2+ i\varepsilon)\ln(q-2+i\varepsilon).
$$

The second derivatives of these functions result in Cauchy kernels
$$
f_{\pm\pm}''(q)=\dfrac{1}{q\pm2\pm i\varepsilon}.
$$

Let's find the second derivative $\varepsilon_l''(q) $. In the
expression for $\varepsilon_l''(q)$ we leave only those items,
which contain Kohn singularities and
result in the Friedel oscillations
$$
\varepsilon''_l(q)=-\dfrac{3x_p^2}{2q^2}\dfrac{g_+''(q)-g''_-(q)}
{1-g_0(q)}=-\dfrac{3x_p^2}{16q^3}\dfrac{g_1''(q)-g_2''(q)}{1-g_0(q)}.
\eqno{(2.3)}
$$

Let's return to the integral (1.6) and by means of (2.3) we
present it in the form
$$
V(r)=-\dfrac{1}{r^3}\cdot \dfrac{3Qx_p^2}{16\pi}
\int\limits_{-\infty}^{\infty}\dfrac{[g_1''(q)-g_2''(q)]\sin(k_Frq)}
{q^4\varepsilon_l^2(q)[1-g_0(q)]}+O(e^{-k_{TF}r}),
\quad r\to \infty.
\eqno{(2.4)}
$$

Let's show that in a statical limit ($\omega \to 0$) the expression
$\varepsilon_l(q)$ is real. Really,
according to the results from \cite{LatYush1} the dielectric
permeability is expressed by equality
$$
\varepsilon_l(q)=1+\dfrac{3x_p^2}{4y^2}\cdot
\dfrac{\displaystyle \int\limits_{-1}^{1}\dfrac{(1-t^2)dt}
{(1-i\omega \tau+iqt/y)^2+q^4/(4y^2)}}{1-\dfrac{1}{2}
\displaystyle\int\limits_{-1}^{1}\dfrac{dt}{1-i\omega \tau+iqt/y}}.
$$

We receive  from here in the statical limit
$$
\varepsilon_l(q)=1+\dfrac{3x_p^2}{4}\cdot
\dfrac{\displaystyle \int\limits_{-1}^{1}\dfrac{(1-t^2)dt}
{(iqt+y)^2+q^4/4}}{1-\dfrac{y}{2}\displaystyle\int\limits_{-1}^{1}
\dfrac{dt}{iqt+y}}.
$$

After simple transformations we reduce the previous expression
to the form which does not contain imaginary unit
$$
\varepsilon_l(q)=1+\dfrac{3x_p^2}{4}\cdot
\dfrac{\displaystyle \int\limits_{-1}^{1}\dfrac{(1-t^2)
(q^4/3+y^2-q^2t^2)dt}
{(q^4/4+y^2-q^2t^2)^2+4q^2y^2t^2}}{1-\dfrac{y^2}{2}
\displaystyle\int\limits_{-1}^{1}
\dfrac{dt}{q^2t^2+y^2}}.
$$

In the expression $g_1''(q)-g_2''(q)$ from the equality (2.4) we will leave the terms containing Kohn singularities and
leading to Friedel oscillations again. As  the result we receive
$$
g_1''(q)-g_2''(q)=
$$
$$
=\dfrac{q-2+i\varepsilon}{q+2-i\varepsilon}-
\dfrac{q+2-i\varepsilon}{q-2+i\varepsilon}
-\dfrac{q+2+i\varepsilon}
{q-2-i\varepsilon}+\dfrac{q-2-i\varepsilon}{q+2+i\varepsilon}.
$$

Now integral (2.4) we will present in the following form
$$
V(r)=-\dfrac{1}{r^3}\cdot \dfrac{3Qx_p^2}{16\pi}
\int\limits_{-\infty}^{\infty}
\dfrac{\sin (k_Frq)}{q^4\varepsilon_l^2(q)[1-g_0(q)]}
\Big[\dfrac{q-2+i\varepsilon}{q+2-i\varepsilon}-
\dfrac{q+2-i\varepsilon}{q-2+i\varepsilon}-\hspace{2.5cm}$$$$\hspace{3.5cm}
-\dfrac{q+2+i\varepsilon}
{q-2-i\varepsilon}+\dfrac{q-2-i\varepsilon}{q+2+i\varepsilon}\Big]
\,dq.
\eqno{(2.5)}
$$

Let's designate now
$$
\varphi(q)=\dfrac{q+(2+i \varepsilon)}{q-(2+i \varepsilon)}+
\dfrac{q+(2-i \varepsilon)}{q-(2-i \varepsilon)}.
$$

It is obvious that other two terms in the square brackets are equal to
$\varphi(-q)$:
$$
\varphi(-q)=\dfrac{q-(2+i \varepsilon)}{q+(2+i \varepsilon)}+
\dfrac{q-(2-i \varepsilon)}{q+(2-i \varepsilon)}.
$$

Now the expression (2.5) we will write in the form
$$
V(r)=\dfrac{1}{r^3}\cdot \dfrac{3Qx_p^2}{16\pi}
\int\limits_{-\infty}^{\infty}
\dfrac{\varphi(q)-\varphi(-q)}
{q^4\varepsilon_l^2(q)[1-g_0(q)]}\sin (k_Frq)\,dq.
$$

This equality can be simplified
$$
V(r)=\dfrac{1}{r^3}\cdot \dfrac{3Qx_p^2}{8\pi}
\int\limits_{-\infty}^{\infty}
\dfrac{\varphi(q)\sin (k_Frq)}
{q^4\varepsilon_l^2(q)[1-g_0(q)]}\,dq.
\eqno{(2.6)}
$$

Let's notice that expression $q^4\varepsilon_l^2(q)$ does not vanish in the point $q=0$. Besides, we will notice that in the linear
approximation of 1$-g_0(q)=1$. Really, we multiplay the subintegral
function from expression for $g_0(q)$ by fraction, numerator and
denominator of which is the expression, conjugating to the denominator
from $g_0 (q) $, we receive:
$$
g_0(q)=\dfrac{i\varepsilon}{2}\int\limits_{-1}^{1}
\dfrac{(qt+iy)\,dt}{q^2t^2+\varepsilon^2}=$$$$=\dfrac{iyq}{2}
\int\limits_{-1}^{1}\dfrac{tdt}{q^2t^2+\varepsilon^2}+
\dfrac{\varepsilon^2}{2}\int\limits_{-1}^{1}\dfrac{dt}{q^2t^2+
\varepsilon^2}=O(\varepsilon^2)\qquad (\varepsilon\to 0),
$$
as it was required to show.

Now the integral from (2.6) we will write down in an explicit form
$$
V(r)=
\dfrac{1}{r^3}\cdot \dfrac{3Qx_p^2}{8}\dfrac{1}{2\pi i}
\int\limits_{-\infty}^{\infty}\Big[
\dfrac{q+(2+i \varepsilon)}{q-(2+i\varepsilon)}+
\dfrac{q+(2-i\varepsilon)}{q-(2-i\varepsilon)}\Big]
\dfrac{e^{ik_Frq}-e^{-ik_Frq}}{q^4\varepsilon_l^2(q)}\,dq.
$$

In last integral subintegral expression has two
singular Cauchy kernels, equal to infinity  in the complex
conjugate points $q=2\pm i \varepsilon $.
Integration
near the point $q=2$ gives the basic contribution to the expression for potential.
For the approximate calculus of last integral we use
the method stated, for example, in the monograph \cite {Harrison}.

For this purpose we calculate continuous functions from subintegral expression
in the point $q=2$:
$$
\Big[q^4\varepsilon_l(q)\Big]\Bigg|_{q=2}=16\Big(1+\dfrac{3}{8}x_p^2\Big)^2.
$$

We receive that
$$
V(r)=\dfrac{1}{r^3}\dfrac{3Qx_p^2}{2(8+3x_p^2)^2}\times
$$
$$
\times
\int\limits_{-\infty}^{\infty}\Big[
\dfrac{q+(2+i \varepsilon)}{q-(2+i\varepsilon)}
+\dfrac{q+(2-i\varepsilon)}{q-(2-i\varepsilon)}\Big]
\Big[e^{ik_Frq}-e^{-ik_Frq}\Big]\,dq.
$$

The integral from the previous equality is equal to the sum of
residues relatively the simple
poles in points $q=2\pm i \varepsilon $:
$$
V(r)=
\dfrac{1}{r^3}\cdot \dfrac{3Qx_p^2}{(8+3x_p^2)^2}
\Big[(2+i\varepsilon)e^{ik_Fr(2+i\varepsilon)}+(2-i\varepsilon)
e^{-ik_Fr(2-i\varepsilon)}\Big].
$$

From here we obtain

$$
V(r)=\dfrac{1}{r^3}\cdot \dfrac{6Qx_p^2e^{-k_Fr\varepsilon}}{(8+3x_p^2)^2}
\Big[2\cos 2k_Fr-\varepsilon\sin 2k_Fr\Big].
$$

Finally we get
$$
V(r)=A\dfrac{e^{-k_Fr\varepsilon}}{r^3}\cos(2k_Fr+\varphi),
$$
or
$$
V(r)=A\dfrac{e^{-k_Fry}}{r^3}\cos(2k_Fr+\dfrac{y}{2}),
\eqno{(2.9)}
$$

Here
$$
\varphi={\rm arctg}\,\dfrac{\varepsilon}{2}=\dfrac{\varepsilon}{2}=
\dfrac{y}{2},
$$
$$
A=\dfrac{12Qx_p^2}{(8+3x_p^2)^2}.
$$

The expression (2.9) can be rewritten in dimensional parameters
$$
V(r)=A\dfrac{e^{-\dfrac{\nu}{k_F}\,
\displaystyle{r}}}{r^3}\cos\Big(2k_F\,r+\dfrac{\nu}
{2k_Fv_F}\Big).
$$

From this formula at $\nu=0$ the classical result for
collisionless plasma is received  (see, for example, \cite {Harrison})
$$
V(r)=\dfrac{A}{r^3}\cos(2k_F\,r).
$$

Let's designate through $R=k_Fr$ dimensionless length, and introduce
dimensionless frequency of collisions
$y=\varepsilon=\dfrac{\nu}{k_Fv_F}$. Then the formula (2.9) for
potential will be written in the form
$$
V(R)=A\,k_F^3\,\dfrac{e^{-Ry}}{{R^3}}\cos(2R+\dfrac{y}{2}).
\eqno{(2.10)}
$$

Graphical study of Friedel oscillations we will carry out for
the case, when $Ak_F^3=10^5$ (fig. 1 - fig. 3 see).
From fig. 3 it is visible that with growth of
frequencies of electron collisions the amplitude of Friedel oscillations
decreases.

\begin{center}\bf
  4. Conclusion
\end{center}

In the present work we use
the expression for dielectric permeability of quantum
degenerate collisional  plasma found by the authors earlier \cite {LatYush1}. With the
help of it the research of influence of collisions on
Friedel oscillations is carried out for the first time.

It is shown that presence of collisions in
plasma results in exponential decrease of amplitude
of Friedel oscillations and to inverse shift of the phase oscillations.
The logarithmic decrement of decrease in linear approximation is equal to
$r=\dfrac{v_F}{\nu}$, and the inverse phase shift is equal to
$\dfrac{\nu}{2k_Fv_F}=\dfrac{1}{2k_Fl}$, where $l$ is the mean free path
of electrons.

\begin{figure}[h]
\begin{flushleft}
\includegraphics[width=17.0cm, height=10cm]{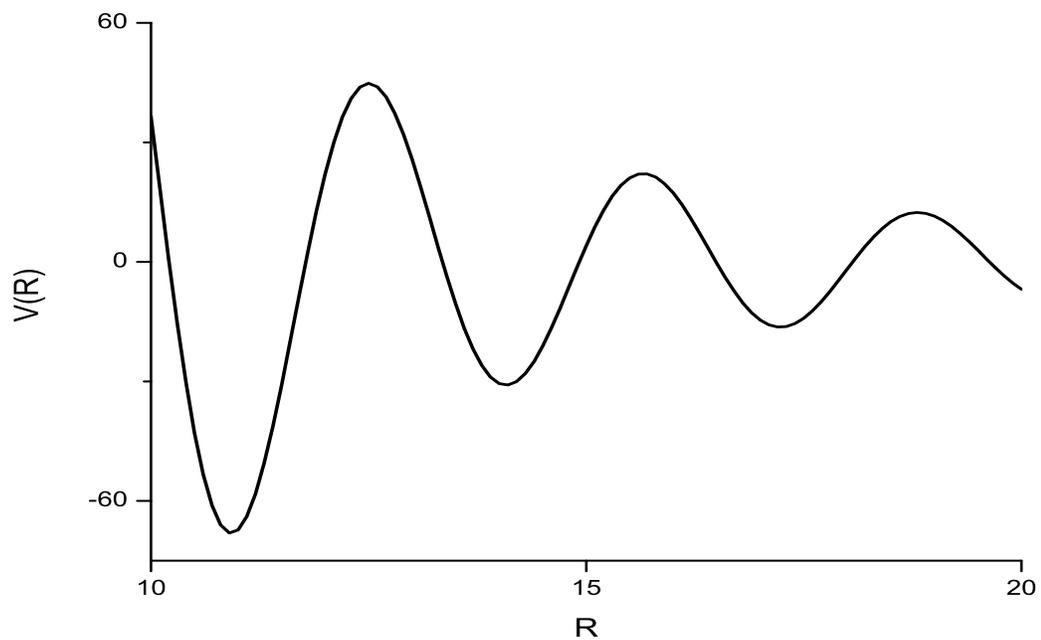}
\caption{Friedel oscillations in the case $y=10^{-2}$, $10<R<20$.
}
\end{flushleft}
\end{figure}

\begin{figure}[t]
\begin{flushleft}
\includegraphics[width=17.0cm, height=9cm]{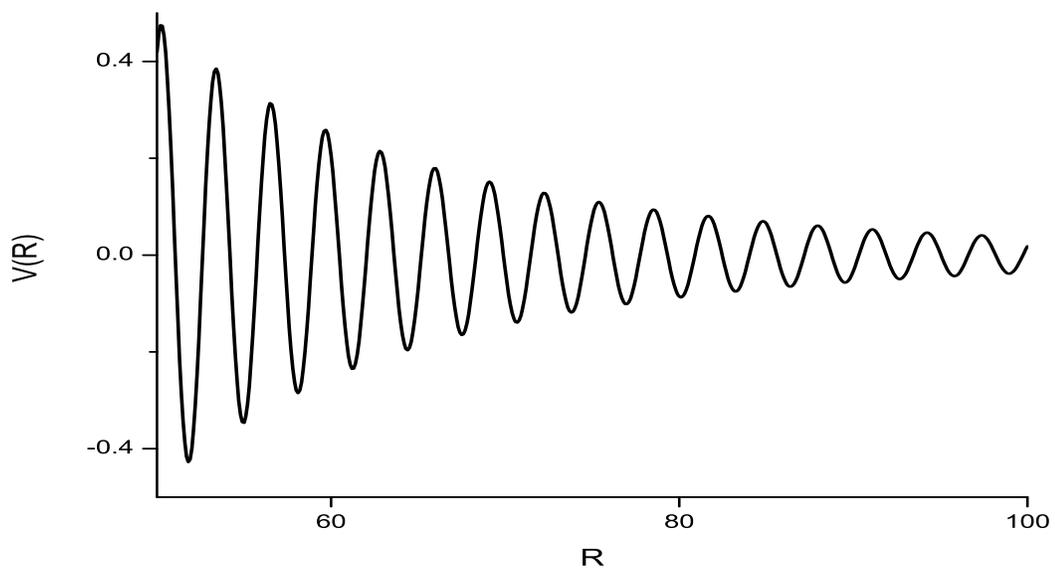}
\caption{Friedel oscillations in the case $y=10^{-2}$, $50<R<100$.
}
\end{flushleft}
\end{figure}
\begin{figure}[b]
\begin{flushleft}
\includegraphics[width=17.0cm, height=9cm]{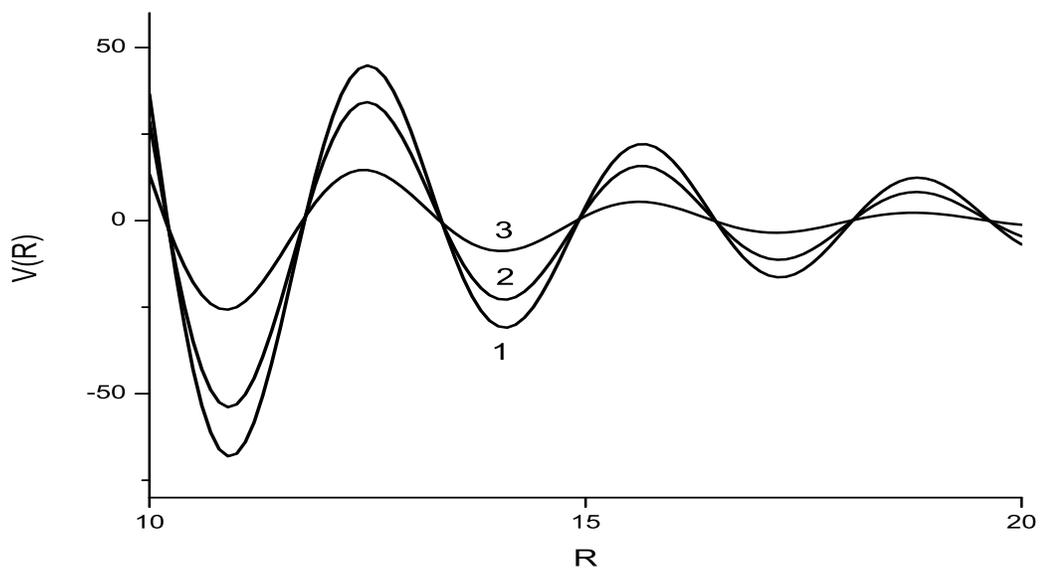}
\caption{Friedel oscillations in the case $y=10^{-2}$,
$10<R<20$. Curves of $1,2,3$ correspond to the values of parameter
 $y, y=10^{-2},
10^{-1.5}, 10^{-1}.$}
\end{flushleft}
\end{figure}

\end{document}